\documentclass[useAMS,usenatbib,usegraphicx]{aastex}
\usepackage{emulateapj5}
\usepackage{times}

\slugcomment{Accepted 17 April 2009}

\shorttitle{Optical rms--flux relation in X-ray binaries}
\shortauthors{P. Gandhi}

\def\rxte{{\sl RXTE}}

\def\ultracam{{\sc ultracam}}
\def\p{$\pm$}
\def\ltsim{\mathrel{\hbox{\rlap{\hbox{\lower4pt\hbox{$\sim$}}}\hbox{$<$}}}}
\def\gtsim{\mathrel{\hbox{\rlap{\hbox{\lower4pt\hbox{$\sim$}}}\hbox{$>$}}}}

\def\aap{A\&A}
\def\nat{Nature}

\def\mnras{MNRAS}
\def\apj{ApJ}

\def\apjl{ApJL}

\def\gx339{GX~339--4}
\def\swiftj1753{SWIFT~J1753.5--0129}
\def\xtej1118{XTE~J1118+480}

\def\barf{$\bar{F}$}

\begin{document}

\title{The flux-dependent rms variability of X-ray binaries in the optical}

\author{P. Gandhi}
\affil{RIKEN Cosmic Radiation Lab, 2-1 Hirosawa, Wakoshi, Saitama 351-0198, Japan}

\label{firstpage}

\begin{abstract}
A linear relation between absolute rms variability and flux in X-ray observations of compact accreting sources has recently been identified. Such a relation suggests that X-ray lightcurves are non-linear and composed of a lognormal distribution of fluxes. Here, a first investigation of the {\em optical} rms vs. flux behavior in X-ray binaries is presented. Fast timing data on three binaries in the X-ray low/hard state are examined. These are \xtej1118, \gx339\ and \swiftj1753\ -- all show aperiodic (non-reprocessed) optical fluctuation components. Optical rms amplitude is found to increase with flux in all sources. A linear fit results in a positive offset along the flux axis, for most frequency ranges investigated. The X-ray and optical relation slopes track the source fractional variability amplitudes. This is especially clear in the case of \gx339, which has the largest optical variance of the three targets. Non-linearity is supported in all cases by the fact that flux distributions of the optical lightcurves are better described with a lognormal function than a simple gaussian. Significant scatter around linearity is found in the relation for the two sources with lower optical variability amplitude, though observational biases may well contribute to this. Implications for accretion models are discussed, and the need for long well-sampled optical lightcurves is emphasized.

\end{abstract}

\section{Introduction}

Several recent works have shown that the X-ray timing characteristics of accreting sources including Galactic black hole candidates, neutron stars as well as active galactic nuclei follow a linear relationship between their root-mean-square (rms) variation and average flux over a wide range of timescales (e.g. \citealt{uttley01} [hereafter, U01], \citealt{uttley04}, \citealt{gleissner04} [hereafter, G04]). Such a relation suggests non-stationarity of the lightcurves; specifically, the level of instantaneous variance in a lightcurve is not constant, but is rather linked to longer-term flux averages. \citet{uttley05} [hereafter, U05] point out that these properties imply a lognormal distribution of instantaneous flare strength. This, in turn, can be explained if the variations are composed not by a superposition of independent shots, but instead are the result of accretion rate fluctuations operating at every location in the accreting flow of matter. These fluctuations propagate inwards through the flow, effectively coupling perturbations at all inner radii. Interestingly, a similar \lq rms--flux relation\rq\ has been shown to apply in other sources as well, for instance in a narrow-line Seyfert galaxy \citep{gaskell04} and in solar coronal X-ray and radio flares \citep[][]{zhang07_conference, wang08_solar_radio_rms_flux}. The relation thus seems to provide strong constraints on the fluctuating conditions of hot plasmas universally.

In this Letter, a first investigation of the rms vs. flux behavior of X-ray binaries (XRBs) in the optical is presented. The presence of fast, aperiodic optical activity in some accreting sources has been known since very early days \citep[e.g. ][]{motch82}. Recent work has found intriguing correlations between the X-ray and optical variations in at least three X-ray binaries: \xtej1118\ \citep{kanbach01}, \gx339\ \citep{g08} and \swiftj1753\ \citep{durant08}. The main driver of the rapid optical variability remains unclear, but several pieces of evidence disfavor an origin as re-processing of, or irradiation by primary higher-energy photons. Instead, it is likely that both the optical and X-ray emitting components share some common underlying cause, with exchange of energy between the components creating the complex interplay of timing patterns observed (cf. \citealt{malzac04}). Studying the optical behavior in the rms--flux plane should provide new clues on the underlying source of variability, independent of commonly-used correlation function analyses. The only other investigation of such optical behavior (as far as the author is aware), is for the active galaxy NGC 4151 for which \citet{lyutyi87} discovered a linear rms--flux relationship in the $U$ band. 

\section{Objects and data}

The sources chosen for this study include all three XRBs known to show direct, accretion-driven rapid variability in the optical. The optical data come from the Skinakas/OPTIMA photometer \citep{optima} for \xtej1118, and from the VLT/ULTRACAM camera \citep{ultracam} for \gx339\ and \swiftj1753. They were observed at optical flux levels several magnitudes brighter than in quiescence so the companion stars contribute negligibly in all cases. The most relevant details including used filter, time resolution and observation period are listed in Table~\ref{tab:obs}. 

The corresponding X-ray data are from the \rxte/PCA detector, observed simultaneously with the optical (but see below). The \rxte\ ObsID prefixes are 50407 for \xtej1118\ and 93119 for both the remaining sources, respectively. The counts and background were extracted over 2--44 keV for \xtej1118\ and over the full PCA energy range for the remaining targets, but the results herein are relatively insensitive to the exact range. Fuller details regarding the X-ray data can be found in the relevant papers cited in the previous section. 

The optical and X-ray data used for \xtej1118\ are strictly-simultaneous. For \gx339, \rxte\ monitoring extends a few minutes longer on either side of the optical monitoring; this full 62 min long dataset is used for better statistics. For \swiftj1753, changeable weather resulted in the best datasets being non-simultaneous; the X-ray data are 54 min long and were observed on 2007 Jun 14 UT, while the optical data are from four days later (see Table~\ref{tab:obs}). All sources maintained a low/hard state throughout the optical/X-ray observing periods.

Multiple nights of observation are available for each source. But a single night stood out as having the best optical weather conditions (in terms of wind for \xtej1118\ and cloud cover for the other two) and only these are used herein. It is noted for completeness that using the strictly-simultaneous optical and X-ray datasets with worse weather conditions, i.e. the non-photometric nights of \swiftj1753\ and \gx339, results in similar behavior in terms of the rms--flux relation described below, but with reduced significance. On the other hand, the nights of \xtej1118\ affected by wind show a departure from this behavior; it is speculated that telescope vibrations may cause loss of flux from the optical fibers of the OPTIMA instrument, rendering these data less useful. 

Power spectral densities (PSDs) for the sources in the listed order in Table~\ref{tab:obs} can be found in \citet{spruitkanbach02}, Gandhi et al. (in prep.) and \citet{durant09}, respectively. These have been recomputed here from averages of long lightcurve segments (256 s) of the same datasets used for the rms--flux relation below, and are displayed in Fig.~\ref{fig:powspec}. The standard rms$^2$-normalization has been used, so that integration over positive frequencies results in the fractional lightcurve variance (\citealt{bellonihasinger90}).

\section{Computing rms vs. flux}

To compute the absolute rms ($\sigma$) as a function of flux, I followed the procedure described by G04 (which is largely similar to that described by U01). In short, the lightcurve for each source was split into many segments of equal length ranging over 1--16 s, depending on the frequency range of interest (see below). The mean source flux in each segment was computed and this distribution was divided into a number of equal-size flux bins (\barf). The binning was optimally chosen so that the fraction of bins containing more than an adopted threshold of 20 segments was maximized; only bins above this threshold were retained in the end. The PSD of each segment (rms$^2$-normalized) was computed, as was the expected white noise level. These two quantities were separately averaged, giving an average PSD in each flux bin as well as an average noise level. Finally, the mean level of the PSD over any particular range of frequencies ($\Delta \nu$) was computed, and white noise subtracted from this. This is the excess (above Poisson) mean fractional variance ($\overline{var}$) in each flux bin, per unit frequency. $\sigma$ is then simply $\sqrt{\overline{var} \Delta \nu} \times \bar{F}$. Uncertainties on the mean $\sigma$ value in each bin are computed with usual periodogram statistics \citep{vanderklis89} and error propagation.

Amongst the observations analyzed herein, there exists a wide range of source count rates, lightcurve sampling ($\Delta T$ and $T$ values in Table~\ref{tab:obs}) and intrinsic source variability levels (Fig.~\ref{fig:powspec}). So there is only a restricted selection of frequency ranges which are ideally suited for inter-comparison of the rms--flux relation. A base range of 0.5--5 Hz was found to be a good compromise for all sources in both X-rays and optical, also helping us to probe the broad-band optical noise continuum free from sharp features (e.g. near $\sim$7--10 Hz) for \xtej1118 and from the low-frequency peaks below 0.1 Hz for the other sources; Fig.~\ref{fig:powspec}). The rms--flux behavior over this frequency range is plotted for all sources in Fig.~\ref{fig:rmsflux}. For more complete comparison, the behavior over two more ranges is also plotted where possible: an appropriate \lq low\rq\ range including frequencies lower than 0.5 Hz, and a \lq high\rq\ range extending beyond 5 Hz. The exact limits of these ranges vary individually according to $\Delta T$, $T$ and count rate. The high frequency range for \xtej1118\ was not computed as it may be affected by telescope vibrations due to wind (Fig.~\ref{fig:powspec}; \citealt{spruitkanbach02}). For \swiftj1753, on the other hand, the total lightcurve duration is too short to produce a meaningful fit at low frequencies, while the PSD variance is too low to give enough signal at high frequencies, restricting the analysis to the base frequency range. 

\section{Results}

\begin{figure*}
  \includegraphics[width=8.5cm]{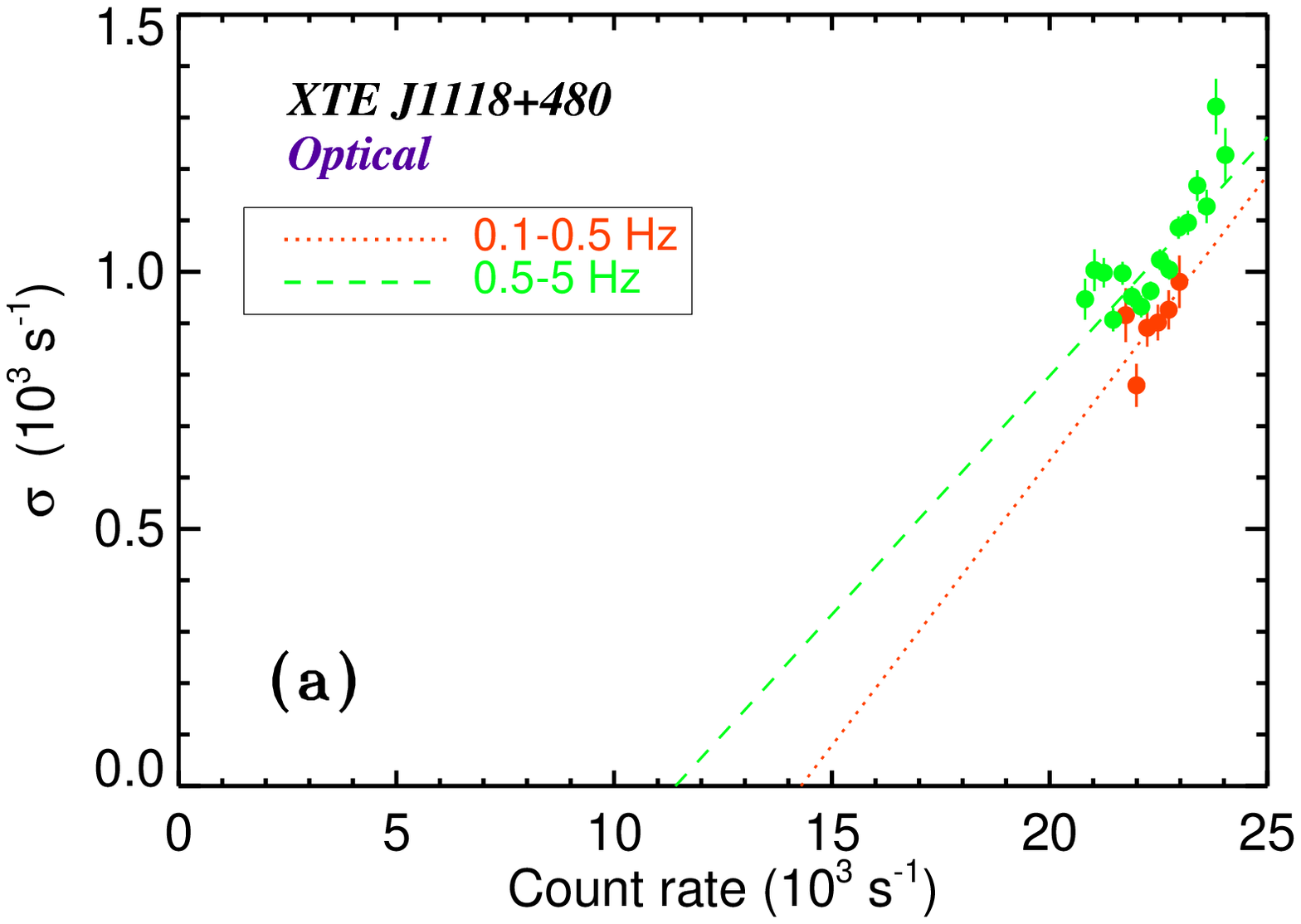}
  \includegraphics[width=8.5cm]{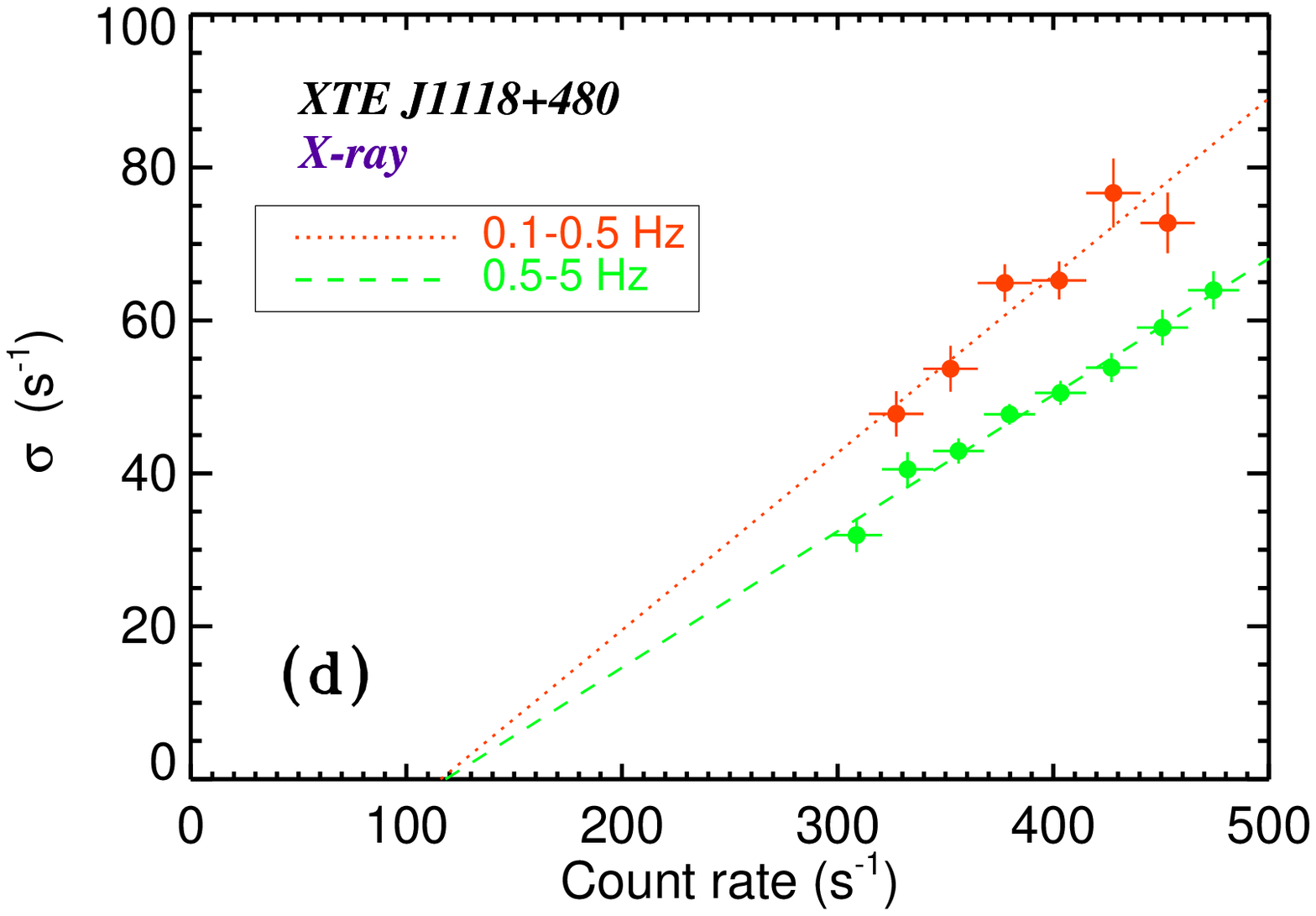}
  \includegraphics[width=8.5cm]{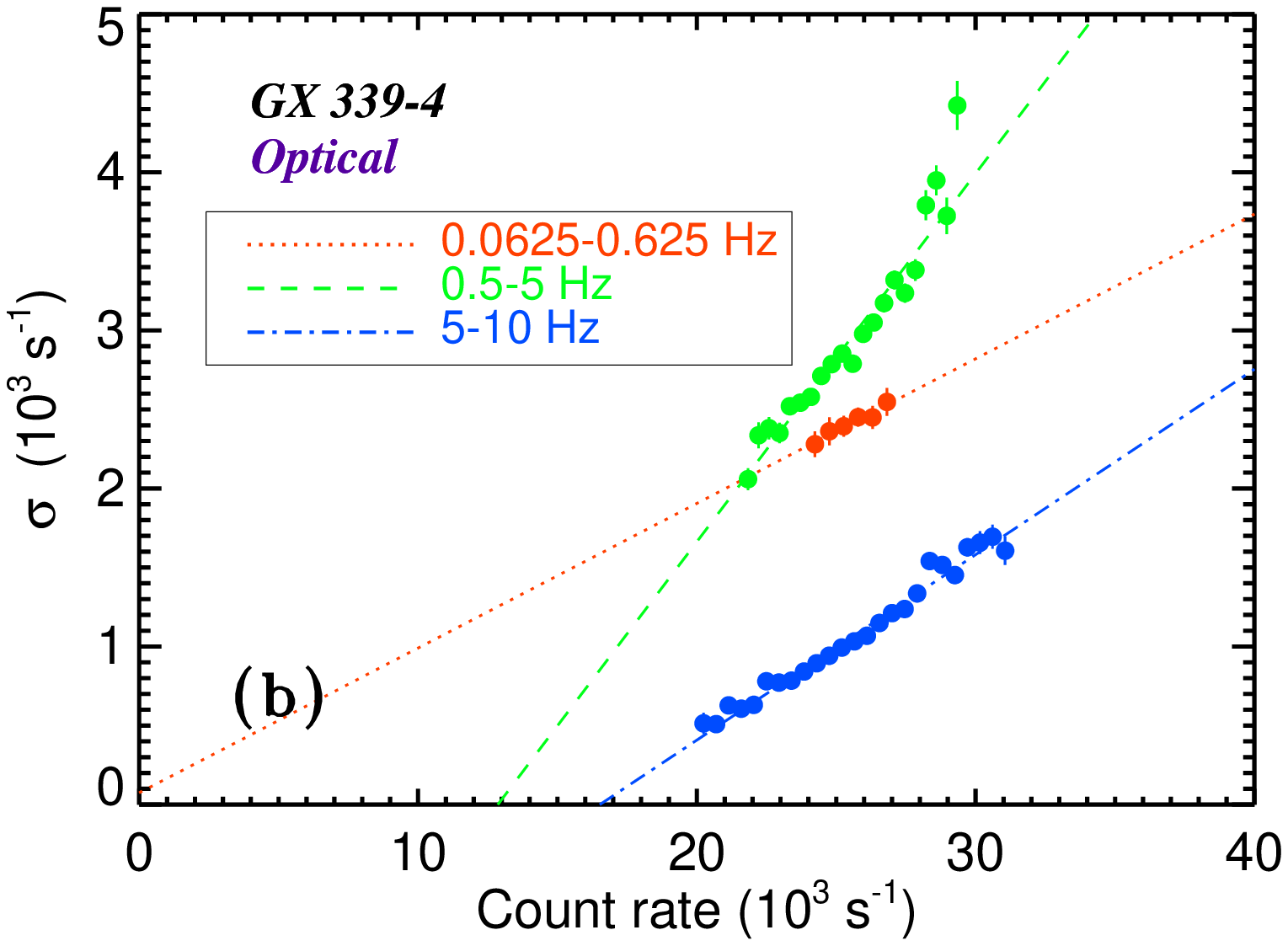}
  \includegraphics[width=8.5cm]{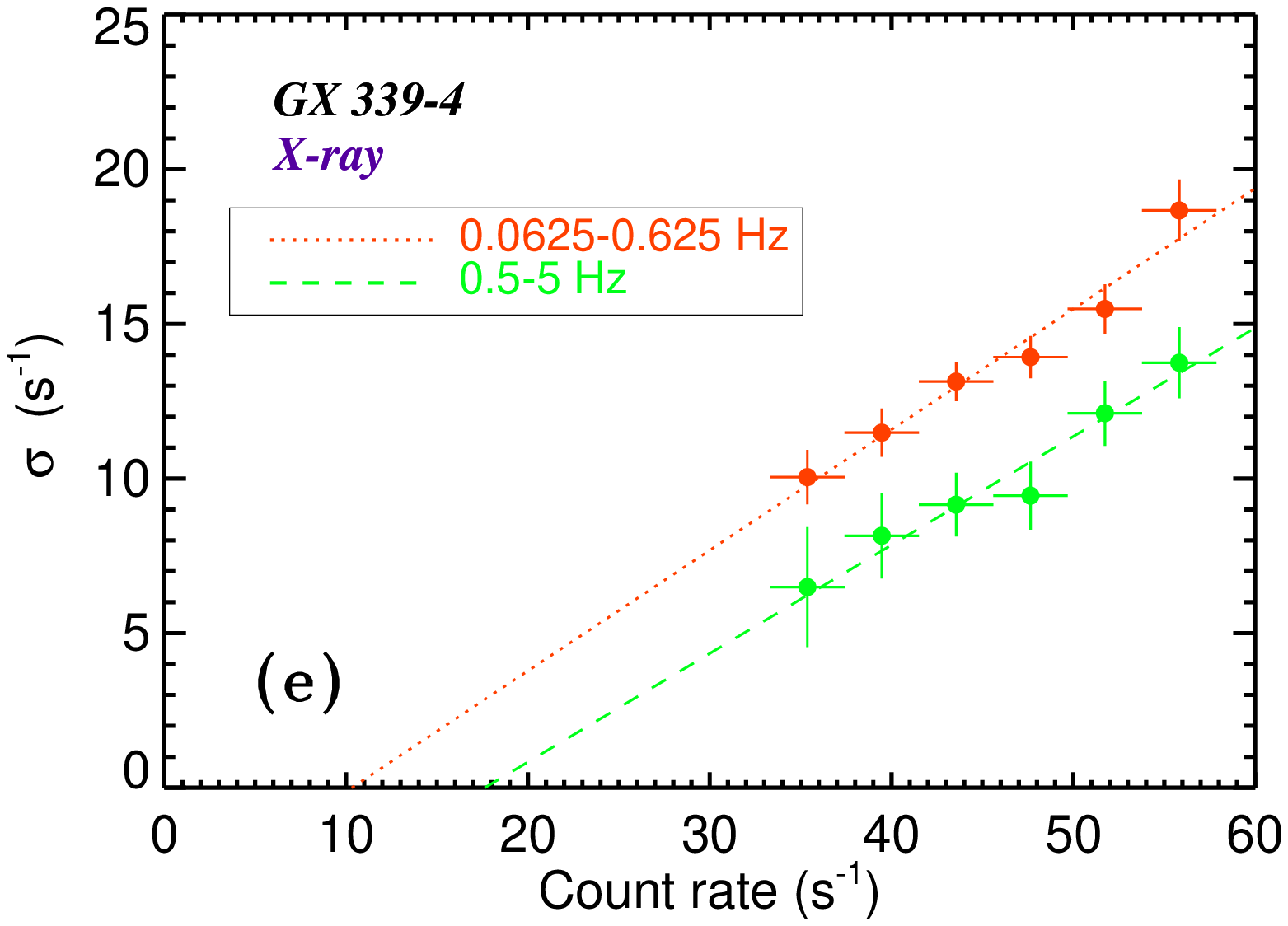}
  \includegraphics[width=8.5cm]{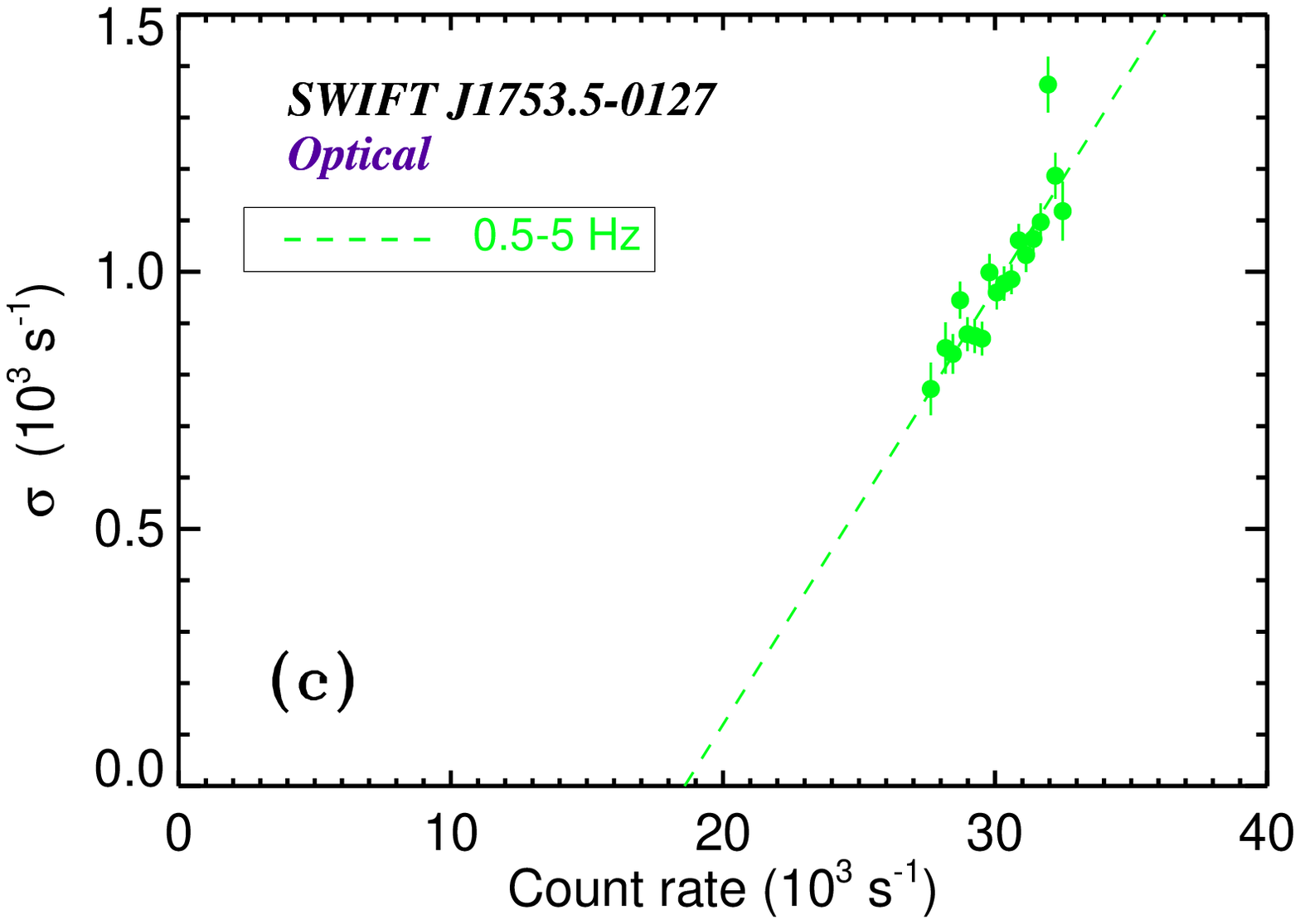}
  \includegraphics[width=8.5cm]{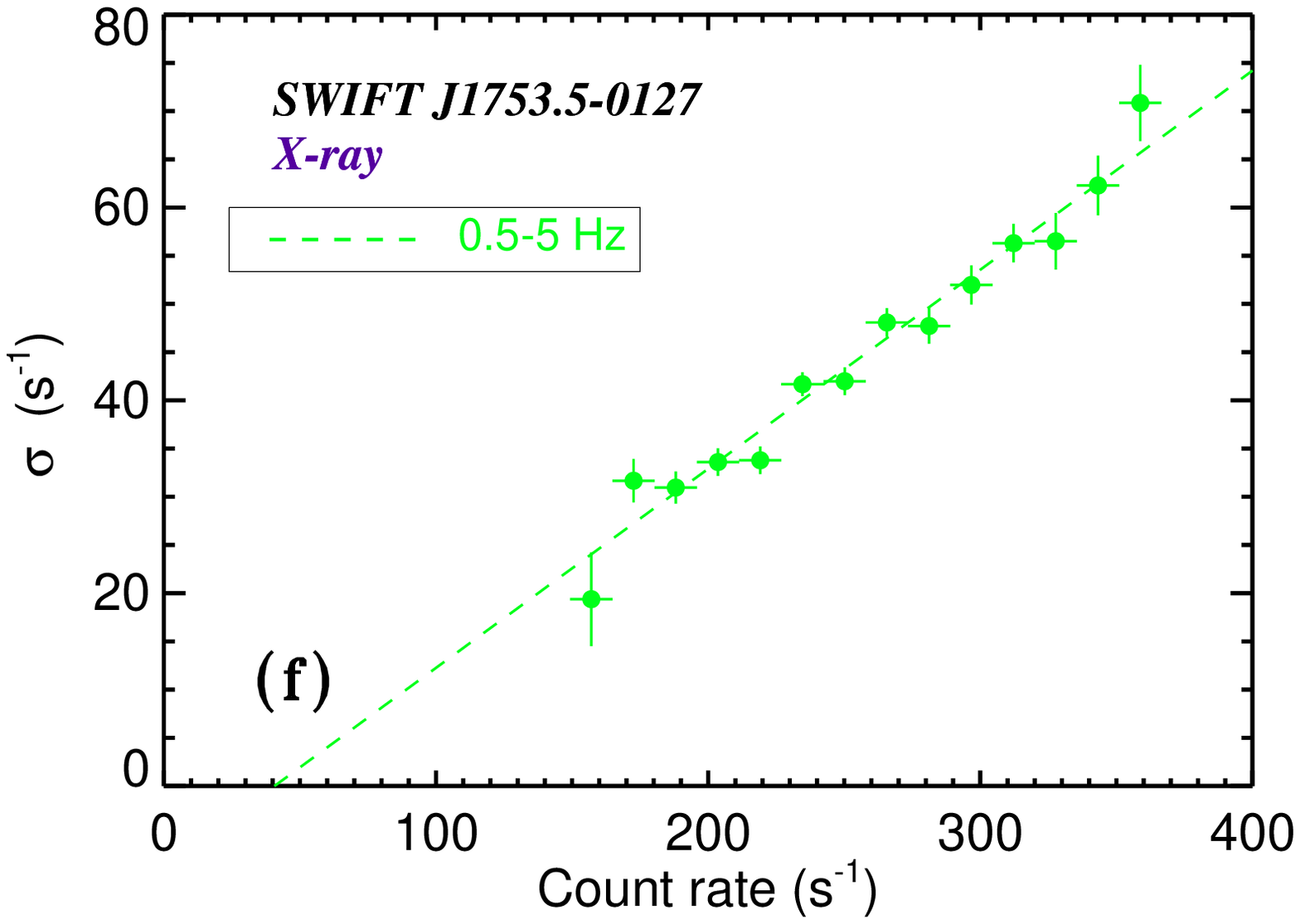}
  \caption{The rms($\sigma$)--flux(\barf) relation in the optical (plots on left) and X-ray (right). For each source, a common base frequency range of 0.5--5 Hz is plotted (green dashed linear fits). Depending on signal:noise, a \lq low\rq\ range (red dotted) and a \lq high\rq\ range (blue dot-dashed) were also computed. The exact ranges are labelled and the corresponding linear fit parameters are listed in Table~\ref{tab:fits}.
\label{fig:rmsflux}}
\end{figure*}

{\em 1.} 
The first important result of Fig.~\ref{fig:rmsflux} is that XRBs do show optical rms increasing with flux over large ranges in frequency. Fitting with the linear parametrization of U01, $\sigma=k($\barf$-C)$, the intercept along the flux axis is generally consistent with a positive offset. These results are listed in Table~\ref{tab:fits}.

{\em 2.} 
Compared to the X-ray $\sigma$--\barf\ relation, the optical relation apparently exhibits more deviations around a linear fit, especially towards the highest and lowest flux bins. This is particularly evident in the case of \xtej1118\ (Fig.~\ref{fig:rmsflux}a) where the scatter in the lowest count-rate bins might even suggest different behavior in this flux regime. Quantifying the linearity of the data with Kendall's rank correlation coefficient ($\tau$, as also used by G04), the optical shows a weaker correlation ($\tau_O$) as opposed to the corresponding X-ray ($\tau_X$) values; these are also listed in Table~\ref{tab:fits}. 

\begin{figure*}
  \includegraphics[width=6cm]{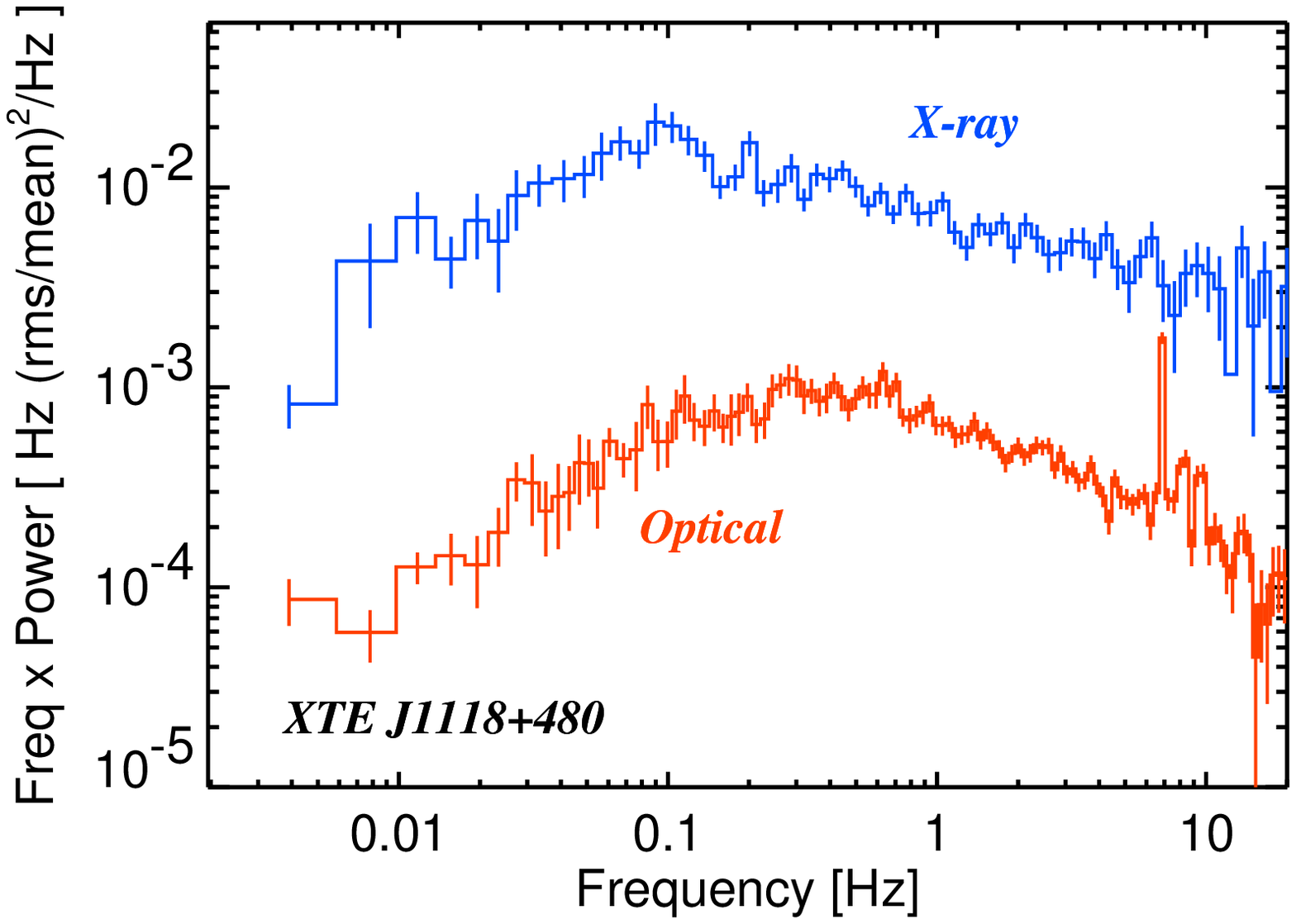}
  \includegraphics[width=6cm]{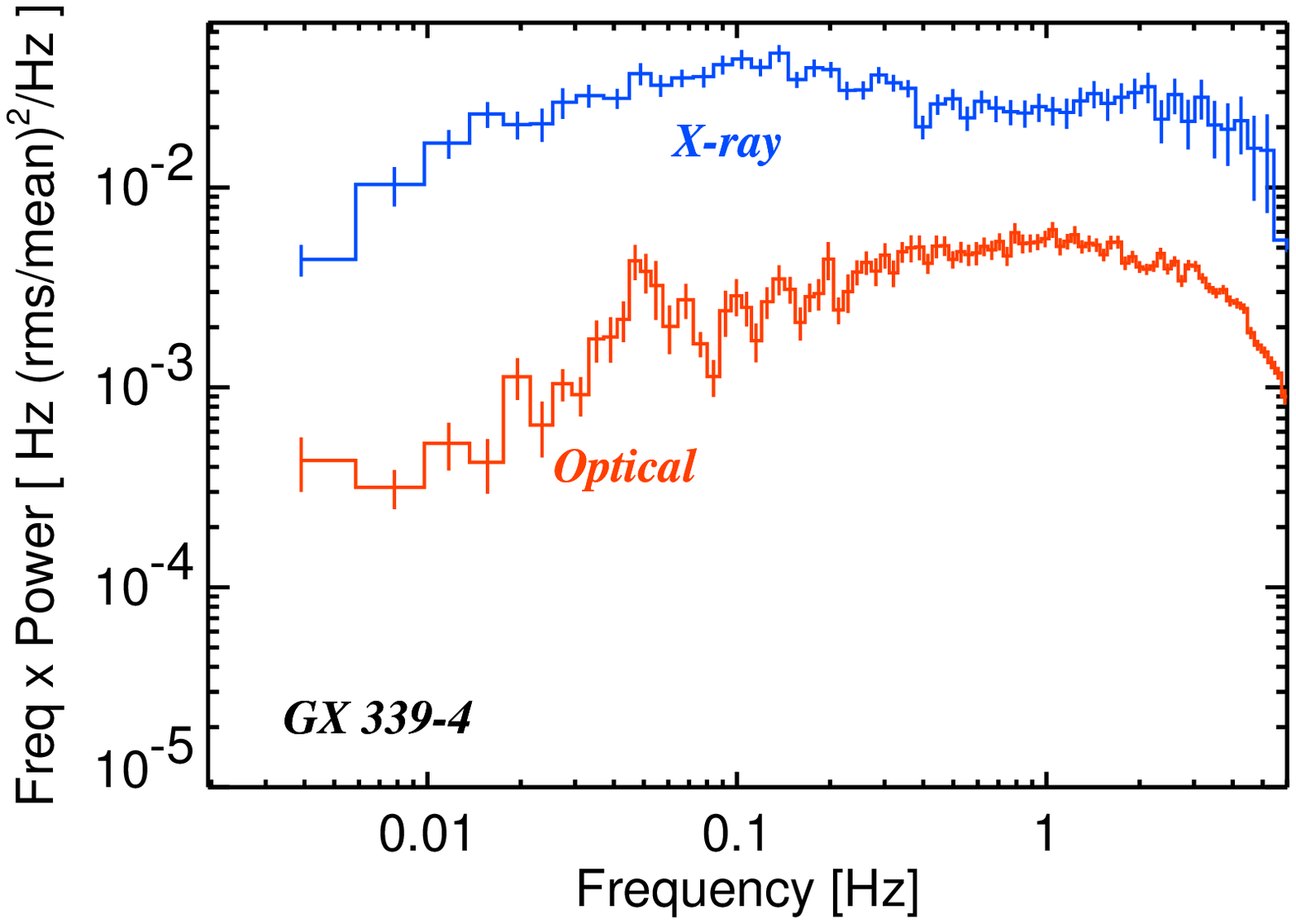}
  \includegraphics[width=6cm]{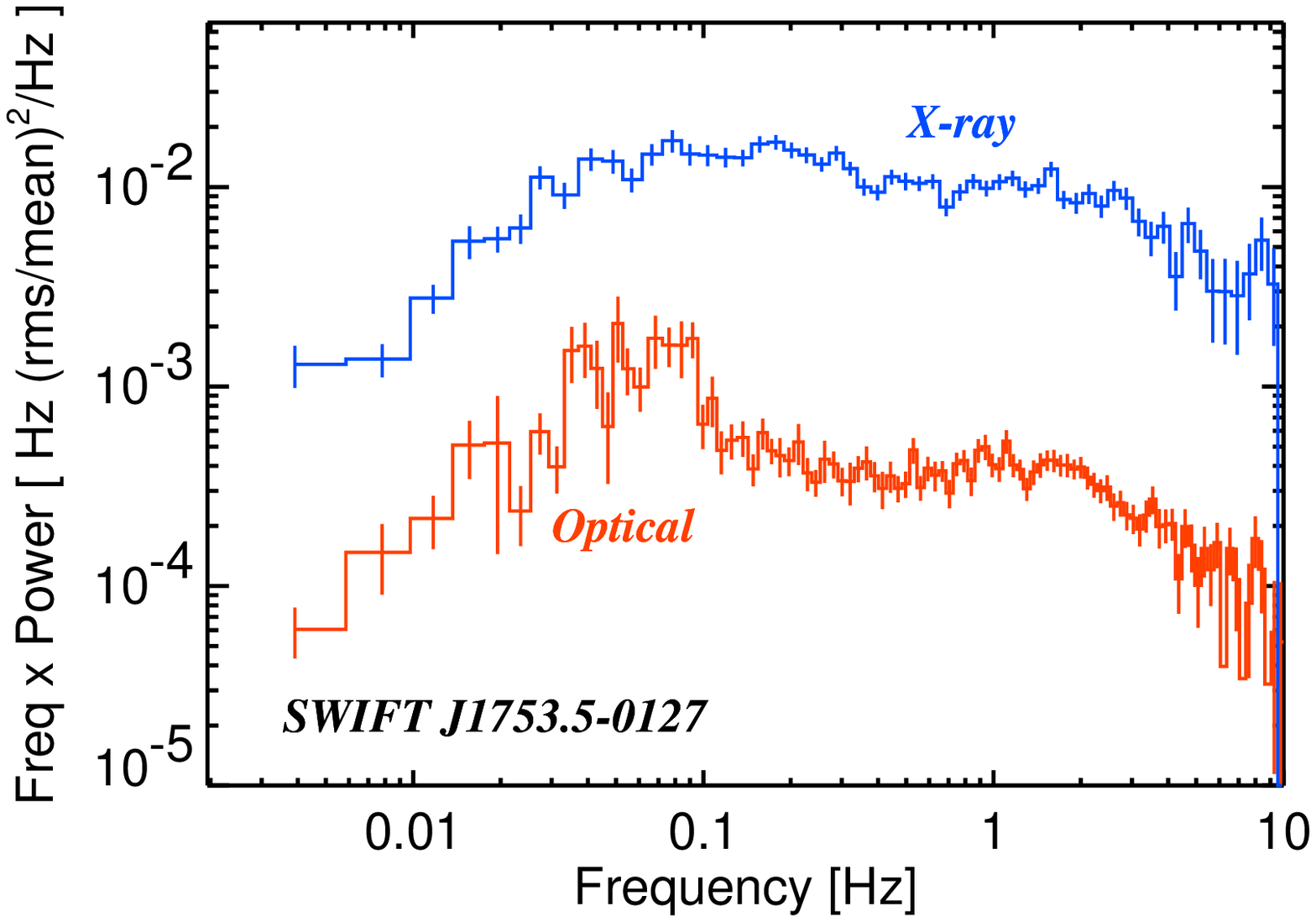}
  \caption{Optical (red) and X-ray (blue) power spectral densities (PSDs) of each source. The PSDs are normalized so that their integral gives the fractional variance of the lightcurves over any frequency interval, and are plotted in units of frequency $\times$ Power/Hz so that a horizontal line denotes equal power per unit frequency-decade. The narrow spikes above $\sim$ 7 Hz in the optical PSD of \xtej1118\ may be a result of telescope vibrations in high wind \citep{spruitkanbach02}. The optical peaks below 0.1 Hz for the remaining two sources are real (cf. Gandhi et al. in prep., \citealt{durant09}).
\label{fig:powspec}}
\end{figure*}

{\em 3.} 
\gx339\ shows the tightest linear correlation in the optical with the highest $\tau_O$ value over all frequency ranges investigated (also evident from Fig.~\ref{fig:rmsflux}b). It is also the source with the highest fractional rms variability amplitude, as can clearly be seen by the fact that its optical PSD has higher broad-band noise than the other sources to well beyond 1 Hz (Fig.~\ref{fig:powspec}).

{\em 4.} 
The rms--flux relation generally flattens (i.e. slope $k$ decreases) with frequency. This trend is consistent with the interpretation that $k$ changes in accordance with the rms of the varying lightcurve component over the frequency range of interest (U01). This can be roughly discerned from the PSDs of Fig.~\ref{fig:powspec}. These have been plotted in units of Frequency $\times$ Power spectral density, so a horizontal line represents equal variability amplitude per decade. Since higher frequencies typically have a lower rms amplitude, the corresponding $k$ values are smaller. Furthermore, the lower optical PSD variance (as compared to X-rays) is manifested in the fact that $k_O<k_X$ (Table~\ref{tab:fits}). Finally, the only case in which the relation steepens with frequency is \gx339, for which $k_O^{\rm low}<k_O^{0.5-5}$; again, this is consistent with the steepening of the optical PSD seen in this case.

\begin{table*}
{\scriptsize
  \begin{center}
  \caption{Optical observations log
    \label{tab:obs}}
    \begin{tabular}{lcccccccr}
      \hline
    Source             &  UT Date    &  Telescope/Instrument  & $\Delta \lambda$ (filter) & $\Delta T$   &    $T$   & $\left( \chi^2/dof\right)_{\rm LN}$  & $\left( \chi^2/dof\right)_{\rm N}$\\
                       &             &                        &    \AA                    &     ms       &     min  &           &       \\
     (1)               &  (2)        &        (3)             &    (4)                    &     (5)      &    (6)   &  (7)      &   (8) \\
    \hline
    \xtej1118          & 2000.07.05  &   Skinakas 1.3m/OPTIMA &     4500--9500 (--)       &      3       &    40    & 79.8/63  & 2587.5/64 \\
    \gx339             & 2007.06.18  &   VLT/ULTRACAM         &     5500--7000 ($r'$)     &     50       &    47    & 76.8/38  & 2712.9/39 \\
    \swiftj1753        & 2007.06.18  &   VLT/ULTRACAM         &     5500--7000 ($r'$)     &     39       &    31    & 28.7/20  &  101.5/21 \\
    \hline
    \end{tabular}

Col. (4) lists the approximate wavelength coverage (and filter name in brackets); In Col. (5), $\Delta T$ refers to the best time resolution available in the optical, in milli-seconds; Col. (6) lists the approximate full length of the lightcurves used in the analysis, in minutes. Cols. (7) and (8) report the chi-square / degrees-of-freedom for the lognormal (\lq LN\rq) and simple normal (\lq N\rq) histogram fits to the distribution of lightcurve fluxes (see section \ref{sec:discussion} and Fig.~\ref{fig:lognormal}).
  \end{center}
}
\end{table*}

\begin{table*}
{\scriptsize
  \begin{center}
  \caption{rms--flux relation fits to the relation $\sigma$=$k$(\barf--$C$)
    \label{tab:fits}}
    \begin{tabular}{l|ccc|ccc|ccr}
      \hline
\underline{Source} &  \multicolumn{9}{c}{\underline{Optical}} \\
           &  $k_O^{\rm low}$ & $C_O^{\rm low}$ & $\tau_O^{\rm low}$ & $k_O^{0.5-5}$  & $C_O^{0.5-5}$ & $\tau_O^{0.5-5}$ & $k_O^{\rm high}$ & $C_O^{\rm high}$ & $\tau_O^{\rm high}$ \\
      (1)  &  (2)             &  (3)            &   (4)              &     (5)        &    (6)        &     (7)          &   (8)          &     (9)        &     (10)  \\
    \hline	       				    	
    \xtej1118   &  0.11\p0.05    &   14.29\p11.8  & 0.60 & 0.09\p0.01   &   11.41\p2.5 & 0.67 &--            &     --         & -- \\
    \gx339      &  0.09\p0.04    &  --0.82\p11.3  & 0.87 & 0.23\p0.01   &   12.86\p1.1 & 0.95 & 0.12\p0.01   &   16.52\p0.9   & 0.94 \\ 
    SWT~J1753.5--0127 &  --            &      --        &    --     & 0.08\p0.01   &   18.59\p3.16& 0.82 & --   & --  &  --  \\
    \hline	       				    	
\underline{Source} &  \multicolumn{9}{c}{\underline{X-ray}} \\
           &     $k_X^{\rm low}$      &     $C_X^{\rm low}$  &  $\tau_X^{\rm low}$  &     $k_X^{0.5-5}$    &     $C_X^{0.5-5}$    &   $\tau_X^{0.5-5}$ &     $k_X^{\rm high}$      &     $C_X^{\rm high}$    &   $\tau_X^{\rm high}$   \\
    \hline	       				    	
    \xtej1118   &  0.23\p0.04    & 115.99\p77.8   &  0.87 & 0.18\p0.02   & 118.2\p47.0  & 1     & --            & --         & --   \\
    \gx339      &  0.39\p0.07    & 10.32\p8.8     &  1    & 0.35\p0.09   & 17.64\p13.2   & 1        & --            & --         & -- \\
    SWT~J1753.5--0127 &     --          &    --          &    -- & 0.21\p0.01   & 41.67\p17.2  & 0.96     & --            & --         & -- \\
    \hline
    \end{tabular}

The slope ($k$) and intercept ($C$) linear fit parameters are stated for three frequency ranges. The first is a \lq low\rq\ range including frequencies below 0.5 Hz (Cols. 2 and 3) along with Kendall's rank correlation coefficient ($\tau$) as a measure of linearity between $\sigma$ and \barf\ (Col. 4). This is followed by the corresponding values for the range of 0.5--5 Hz (Cols. 5, 6 and 7) and a \lq high\rq\ range at greater frequencies (Cols. 8, 9 and 10). See Fig.~\ref{fig:rmsflux} for corresponding plots and details. The fits to the optical are in the first three rows and are denoted by subscript $O$. The corresponding parameters for the X-ray fits are below these, with subscript $X$. The units of $C$ are cts s$^{-1}$, with the optical values ($C_O$) being normalized to 1000 cts s$^{-1}$ for better legibility. Errors quoted are for $\Delta \chi^2$=1 on each fit parameter.
  \end{center}
}
\end{table*}

\section{Discussion}
\label{sec:discussion}

A linear-like rms--flux relation is exhibited by XRBs in the optical. The scatter in this relation is apparently larger than in X-rays, but this may well be the result of observational biases. 

For instance, the optical data for \xtej1118\ were observed without any filters over the full optical wavelength range (Table~\ref{tab:obs}). The origin of the variable optical emission from this source remains unclear, with several components including the disk, accretion flow and jet likely to contribute (e.g. \citealt{merloni00}, \citealt{esin01}, \citealt{chaty03}, \citealt{yuan05}, \citealt{reis09}). This could allow for the emission from distinct components to overlap and introduce significant scatter in any underlying strong correlation. On the other hand, it is to be noted that the X-ray data have lower signal:noise than the optical. So the larger error bars could well be hiding more scatter in X-rays, because the $\tau$ statistic does not account for measurement uncertainties. 

Furthermore, optical timing observations are sensitive to prevailing weather conditions. In the case of \xtej1118, at least, there is some evidence of the wind contaminating the variability power \citep{spruitkanbach02}, though only the night least affected by this bias has been chosen here. Finally, the fact that the overall source variability rms levels are much lower in the optical is an important limitation. The impact of this is that the X-ray relations in Fig.~\ref{fig:rmsflux} cover a wider dynamic range on the x-axis (\barf) and can hence be better determined than the corresponding optical ones. 

With these caveats in mind, the detection of a linear rms--flux relation is important because it provides new constraints for any model attempting to explain the underlying mechanism of the rapid variability. Uttley and collaborators have ruled out additive shot and self-organized criticality models, as these are unable to easily reproduce a linear $\sigma$--\barf\ relation in X-rays over a wide range of timescales. From observations of the accreting neutron star SAX~J1808.4--3658 and correlating the source rms with the strength of high-frequency coherent pulsations, \citet{uttley04} suggests that the rms--flux relation must originate within the accretion flow itself, and not in independent coronal magnetic flares. So-called \lq propagating perturbation\rq\ models are instead favored \citep[][]{lyubarskii97, misra00, king04, titarchuk07} in which accretion rate variations at any given radius in the accretion flow propagate towards smaller radii to modulate the radiative energy release in all inner regions. This naturally couples the variability over a wide range of radii (hence, timescales) and results in a flux-dependent rms amplitude. U05 have shown that this also implies that X-ray lightcurves must be intrinsically non-linear; specifically, the X-ray flux distribution ought to be have a lognormal distribution.

\begin{figure*}
  \includegraphics[width=8cm]{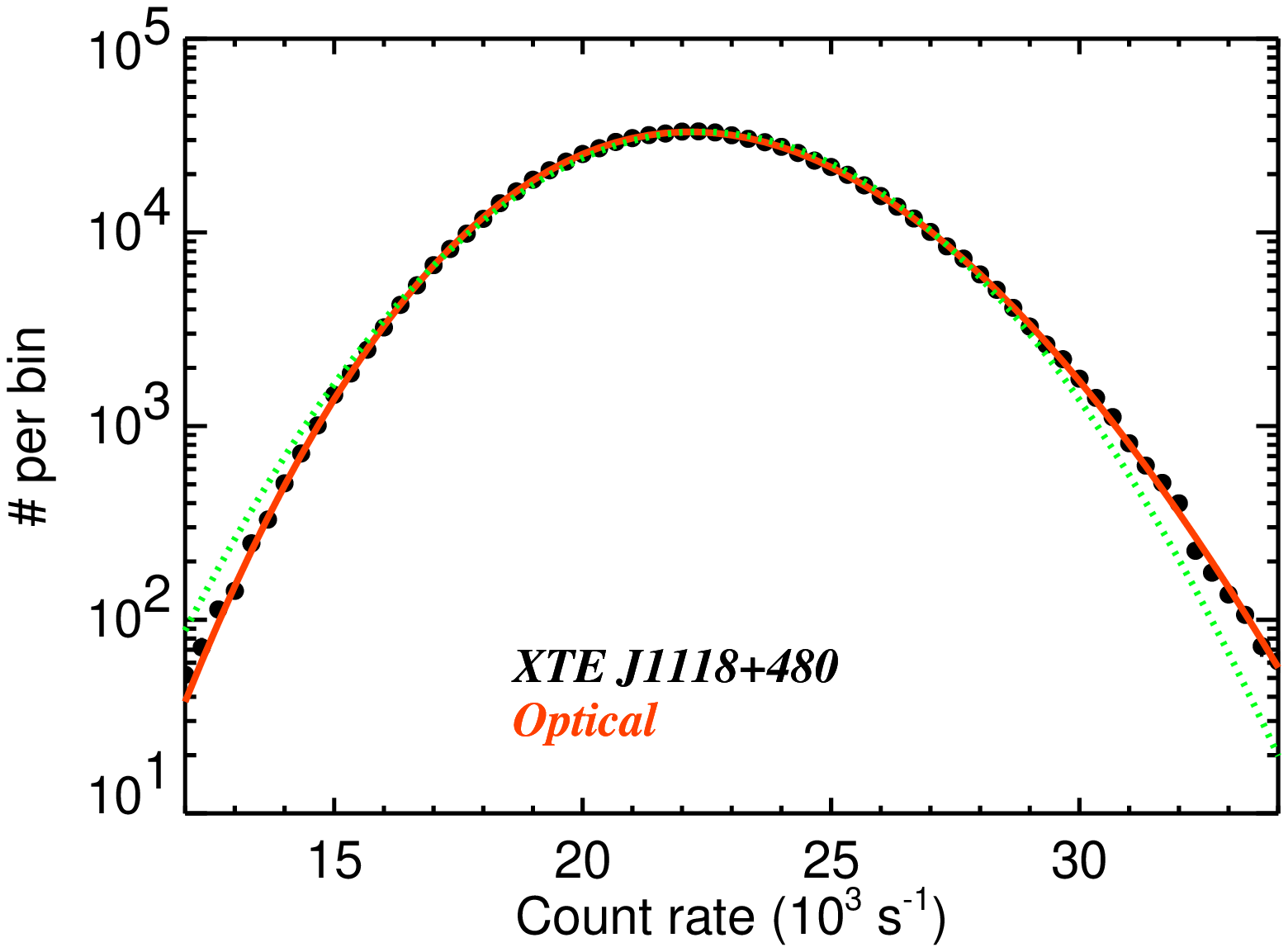}
  \includegraphics[width=8cm]{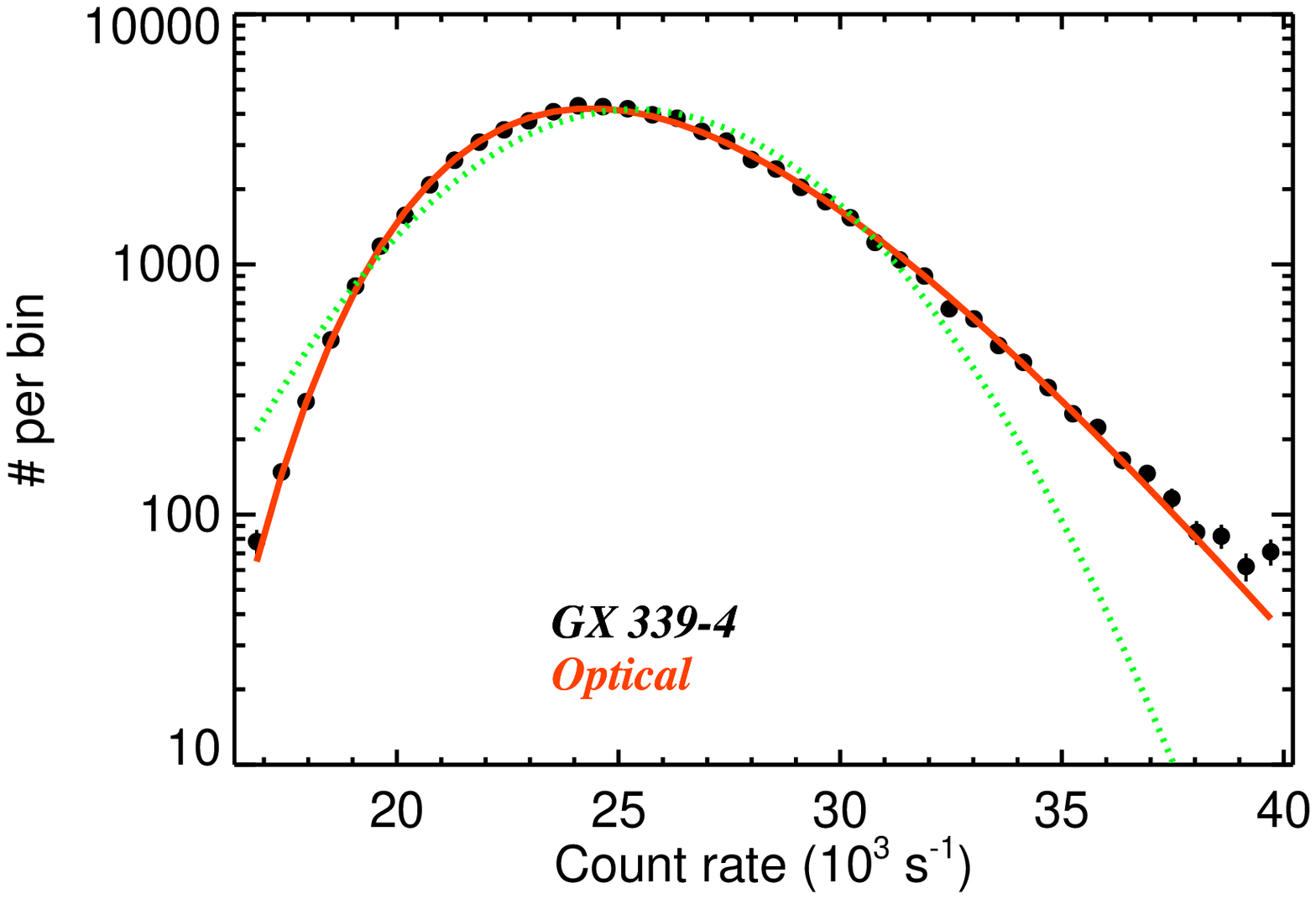}
  \includegraphics[width=8cm]{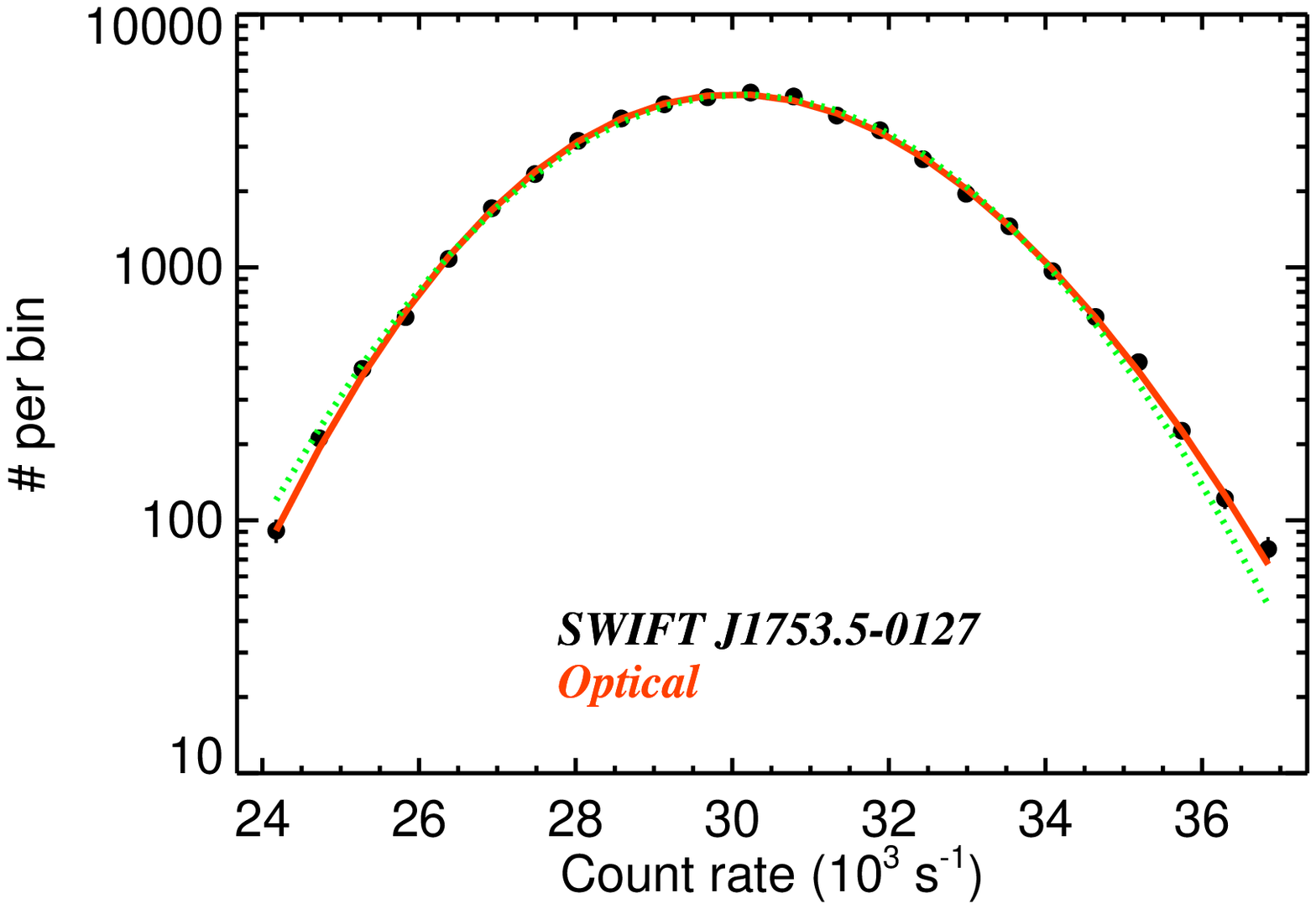}
  \caption{Binned frequency distribution of mean optical source flux. In each case, a lognormal distribution fit is shown by the red curve, while the dotted green curve is the best symmetric gaussian fit. All cases are statistically better-described by lognormals. The y-axes are plotted with a logarithmic scaling in order to accentuate the differences between the curves.
\label{fig:lognormal}}
\end{figure*}

The present work now extends these results to the variable {\em optical} flux of XRBs. We can then also test the above hypothesis of non-linearity by plotting the distribution of optical lightcurve fluxes. These are shown in Fig.~\ref{fig:lognormal} for all three XRBs. In each case, we fit a symmetric gaussian as well as a three-parameter lognormal model (cf. Eq.~3 of U05). The number of flux measurements per bin ($N$) was required to be at least 50 for the fit, with the bin error assigned as $\sqrt{N}$. 

\gx339\ clearly exhibits a lognormal distribution with a very significant tail and even some apparent excesses above the fit in the highest flux bins. The other sources are also better-described as lognormals, though the difference with respect to a simple gaussian fit is more modest, at least to the eye. The best-fit $\chi^2$ values for all fits are listed in Table~\ref{tab:obs}. For completeness, it is noted that the lognormal count-rate offset parameter ($\tau$ in Eq. 3 and section 4.1 of U05) was found to be non-zero in each case; again, this could be best-constrained in the case of \gx339, where it matches the range of values for the rms--flux relation offset ($C_O$).

As mentioned by U05, lognormal fits may not be formally acceptable due to weak non-stationarity of the lightcurves and other effects; this may explain some of the excess in the $\chi^2$ values. But, as can be inferred from Fig.~\ref{fig:lognormal}, the $\chi^2$ residuals for \gx339\ are largely dominated by the highest three (aforementioned) flux bins alone. So \gx339\ is the source with the highest variability (Fig.~\ref{fig:powspec}), and it displays the tightest rms--flux relation as well as the \lq most\rq\ non-linearity. Evidence for non-linearity is present in the other two objects, but is less striking; both exhibit lower broad-band noise. This behavior is exactly as predicted by U05, but now applied to the optical lightcurves. 

Whatever the physical mechanism of variable optical emission may be (synchrotron from the base of a jet is one possible source often discussed in the literature), it should be the result of modulations from an ensemble of coupled perturbations. As compared to auto- and cross-correlation function analyses, which are most sensitive to coherent variations with the shortest delays, the rms--flux relation reveals underlying connections over much wider timescales. The new results presented herein are based on only the handful of relatively-short optical observations which happened to capture the sources at differing stages of their X-ray low/hard states. For instance, \gx339\ was observed at the onset of this state with a relatively faint optical counterpart, while \xtej1118\ was seen in its optically-brightest state. The relative fluxes of the various accretion components are bound to change as the sources evolve through their states (e.g. \citealt{g08}, and also the measurement of the X-ray rms--flux relation in several states by G04). Longer lightcurves at multiple optical and X-ray wavelengths, and in various source states, will greatly aid more precise comparisons. 

In the low/hard state, the disk is not the sole contributor to the optical and X-ray fluxes. Hence the coupling of perturbations must span not only many radii, but also various accretion structures, in order for it to be observed in the Comptonized components and the jet \citep[e.g. ][ and references therein]{done07}. On the other hand, \citet{zhang07_conference} stresses the fact that solar flares follow a similar rms--flux relation which, in turn, disfavors propagating perturbation models, because there is no accretion flow in the solar corona. \citeauthor{zhang07_conference} suggests that a large-scale inter-connected magnetic field in which fluctuations are able to rapidly diffuse to all parts of the corona may work instead. More investigation is needed to understand such a model in detail; but in this context, we note that the reservoir model of \citet{malzac04} could prove to be consistent with all the observations. Their study proposes that the strength of the emergent variable power scales in proportion to the total amount of energy stored in some \lq reservoir\rq. Though energy injection in their model is parametrized as additive linear shots, a quasi-linear rms--flux relation is predicted in X-rays; this is a consequence of coupling between the fluctuating energy dissipation rates within the jet and corona. A simple modification of the model to include non-linear shot amplitudes (by taking the exponential of the lightcurves; cf. U05) will result in a similar relation in the optical as well. This could be achieved with minimal changes to the resultant optical/X-ray coherence and cross-correlations functions. A good candidate for the reservoir is a large-scale magnetic field. Cumulative buildup and dissipation of energy within the coronal+jet (azimuthal+poloidal) field should result in the coupled modulations required by the rms--flux relation.

\acknowledgements
PG acknowledges a RIKEN Foreign Postdoctoral Research Fellowship. The \xtej1118\ data were obtained by G. Kanbach and H.C. Spruit, and the author thanks them as well as J. Malzac for access to these. The efforts of the \ultracam\ team and ESO staff for acquisition of the \gx339\ and \swiftj1753\ optical data (as part of ESO program 079.D-0535) are much appreciated. T. Marsh is thanked for making public his \ultracam\ pipeline. \ultracam\ is supported by STFC grant PP/D002370/1. This work was triggered by comments by T. Belloni on earlier work by the author. PG thanks the anonymous referee for helpful comments.

\label{lastpage}
\end{document}